\documentclass[aps, prx, 10pt,
    notitlepage,superscriptaddress,nofootinbib,
    tightenlines,preprintnumbers,twocolumn]{revtex4-2} 
\usepackage[english]{babel}
\usepackage{amsmath,
    amssymb,
    latexsym,
    mathrsfs,
    siunitx,
    mathtools,
    verbatim,
    indentfirst,
    url,
    enumitem,
    graphicx,
    lipsum,
    afterpage,
    epstopdf,
    orcidlink
}
\usepackage[title,titletoc]{appendix}
\usepackage{hyperref}
\hypersetup{
    colorlinks=true,
    linkcolor=blue,
    filecolor=blue,      
    urlcolor=blue,
    citecolor=blue
}

\makeatletter
\def\p@subsection{}
\makeatother

\makeatletter
\def\p@subsubsection{}
\makeatother

\makeatletter
\newcommand\footnoteref[1]{\protected@xdef\@thefnmark{\ref{#1}}\@footnotemark}
\makeatother

\newcommand{\avg}[1]{\ensuremath{\langle#1\rangle}}
\DeclareSIUnit \belm {Bm}

\begin{document}
\title{Experimental observation of multimode quantum phase transitions in a superconducting Bose-Hubbard simulator} 

\author{Claudia Castillo-Moreno\,\orcidlink{0000-0003-4626-2885}}
\email{claudiac@chalmers.se}
\affiliation{Department of Microtechnology and Nanoscience, Chalmers University of Technology, 412 96 Gothenburg, Sweden}
\author{Théo Sépulcre\,\orcidlink{0000-0002-2434-4487}}
\altaffiliation[Present address: ]{Center for Quantum Computing, RIKEN, Wako-shi, Saitama 351-0198, Japan}
\affiliation{Department of Microtechnology and Nanoscience, Chalmers University of Technology, 412 96 Gothenburg, Sweden}
\author{Timo Hillmann\,\orcidlink{0000-0002-1476-0647}}
\affiliation{Department of Microtechnology and Nanoscience, Chalmers University of Technology, 412 96 Gothenburg, Sweden}
\author{Kazi Rafsanjani Amin\,\orcidlink{0000-0002-1596-1424}}
\affiliation{Department of Microtechnology and Nanoscience, Chalmers University of Technology, 412 96 Gothenburg, Sweden}
\author{Mikael Kervinen\,\orcidlink{0000-0002-0164-9173}}
\altaffiliation{Present address: VTT Technical Research Centre of Finland Ltd. Tietotie 3, Espoo 02150, Finland}
\affiliation{Department of Microtechnology and Nanoscience, Chalmers University of Technology, 412 96 Gothenburg, Sweden}

\author{Simone Gasparinetti\,\orcidlink{0000-0002-7238-693X}}
\email{simoneg@chalmers.se}
\affiliation{Department of Microtechnology and Nanoscience, Chalmers University of Technology, 412 96 Gothenburg, Sweden}


\begin{abstract}
The study of phase transitions and critical phenomena arising in quantum driven-dissipative systems, and whether a correspondence can be drawn to their equilibrium counterparts, is a pressing question in contemporary physics. The development of large-scale superconducting circuits provides an experimental platform for these theoretical models. We report an experimental study of a multi-mode dissipative first-order phase transition in a 1D Bose-Hubbard chain consisting of 21 superconducting resonators. This phase transition manifests itself as a simultaneous frequency jump in all resonator modes as the frequency or power of a pump tone is swept.
By measuring the system's emission spectrum through the transition, we characterize the dim-to-bright phase transition and construct the full phase diagram. We further perform time-dependent measurements of the switching between the two phases in the transition region, from which we corroborate the transition line and extract transition times ranging from a few ms up to 143~s. 
Our model, based on single-mode mean-field theory and cross-Kerr interactions, captures the features at moderate pump powers and quantitatively reproduces the transition line. Our results open a new window into non-equilibrium quantum many-body physics and mark a step toward realizing and understanding dissipative phase transitions in the thermodynamic limit using superconducting quantum circuits.
\end{abstract}

{\let\newpage\relax\maketitle}

\section{Introduction}
First-order dissipative phase transitions arise in nonlinear open quantum systems when the steady-state of the system exhibits a discontinuous jump as an external control parameter, such as drive amplitude or frequency, is varied. Across the transition, the system exhibits bistability characterized by the coexistence of more than one steady-state, and a relaxation dynamics slower than the system's intrinsic timescales
\cite{minganti2018, zurek2005, rossini2021, risken1987}. 

Driven-dissipative phase transitions have been studied theoretically~\cite{zhirov2008, stitely2020, caleffi2023, minganti2018, zurek2005, minganti2021, minganti2023, goncalves2025, soriente2018, soriente2021} and experimentally across various platforms, including cold atomic gases~\cite{baumann2010, greiner2002, ferri2021, labouvie2016}, quantum dots \cite{schliemann2003, russell2007, krebs2010,  kessler2012}, and superconducting circuits~\cite{siddiqi2005, krantz2013, weissl2015, fink2017, fitzpatrick2017, mavrogordatos2017, muppalla2018, chen2023, sett2024, pausch2024, beaulieu2025, beaulieu2025a, berdou2023, brookes2021}. These transitions have recently received growing interest due to their potential applications in areas such as quantum sensing~\cite{ fernandez-lorenzo2017, chu2021, dicandia2023, alushi2025, beaulieu2025, petrovnin2024, rota2019} and quantum computation \cite{gravina2023, berdou2023}. In superconducting circuits, first-order dissipative phase transitions have been measured in arrays of Josephson junctions~\cite{siddiqi2005}, nonlinear resonators~\cite{krantz2013, andersen2020, chen2023, beaulieu2025, beaulieu2025a, muppalla2018}, and linear resonators coupled to nonlinear two-level systems~\cite{mavrogordatos2016,  brookes2021, sett2024, berdou2023}. 

Previous experimental studies have focused on measuring how the non-equilibrium dynamics of the transitions scale with different parameters, like the number of particles or couplings \cite{krantz2013, weissl2015, fitzpatrick2017, chen2023}, extracting the Louvillian gap \cite{beaulieu2023} as well as characterizing the hysteresis around the bistable region \cite{beaulieu2023, muppalla2018}. Recently, some works have also extracted the phase diagram of the transition~\cite{petrovnin2024, sett2024}. 
%
%
Despite this progress, the experimental observation of multimode first-order dissipative phase transitions in a Bose-Hubbard metamaterial has remained unexplored. Moreover, this type of transition has been theoretically studied for 2D systems, but discarded for 1D chains~\cite{vicentini2018}. 

In this work, we present clear evidence of a first-order driven-dissipative phase transition in a 1D array of 21 superconducting nonlinear resonators. This type of system, not studied before, approaches the thermodynamic limit. We experimentally observe multimode transitions when only one of the modes is pumped. We characterize the phase diagram of these transitions with both power spectral density (PSD) measurements of the emitted field and time-dependent measurements of the jump rates between dim and bright phases, and find that the two methods give the same transition line. A simplified single-mode excitation model that incorporates self- and cross-Kerr nonlinearities reproduces the observed behavior across pump powers in semi-classical mean-field simulations, using pump power as the only free parameter. 

The structure of the paper is as follows: Section~\ref{section:device_characterization} introduces the device and experimental setup and the spectroscopy of the device at low power, showing the metamaterial mode's distribution. In Section~\ref{section:multi-mode_phase_transition_spectroscopy},  we measure the scattering parameter $\vert S_{21} \vert$, while we sweep the frequency of a pump at a constant power. This measurement gives rise to multimode frequency-dependent phase transitions, which we explain with a single-mode model. In Section~\ref{section:phase_diagram}, we extract the phase diagram of the system based on PSD measurements. In Section~\ref{section:dynamics_close_to_the_phase_transition}, we run time-dependent measurements to obtain the dynamics of the system. We find a good agreement between the transition point obtained from the time-dependent measurements and the phase diagram from the PSD. We end with a summary and outlook in Section~\ref{section:conclussion}.

\section{Device characterization} \label{section:device_characterization}
\begin{figure}
    \centering
    \includegraphics[width=0.48\textwidth]{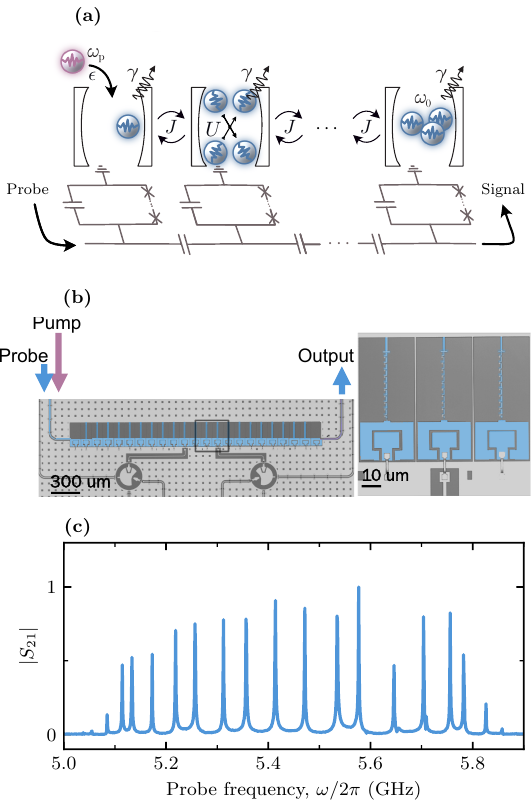}
    \caption{\textbf{Bose-Hubbard chain characterization. (a)} Schematic representation of our experiment, with N cavities with energy $\omega_0$, nonlinearity $U$, jumping rate $J$, and decaying rate $\gamma$. The circuit diagram with LC resonators capacitively coupled is also included. \textbf{(b)} False-color micrograph of the device, including external connections.
    Blue: lumped-element resonator array forming the
    Bose-Hubbard chain. The inset, marked with the black square, shows three resonators with their array of Josephson junctions as the nonlinear inductance.  \textbf{(c)} Transmission of the chain as a function of frequency, with the 21 modes of the system showing up as peaks.}
    \label{fig1}
\end{figure}

Our experiment implements a Bose-Hubbard system described by the Hamiltonian~[Fig.~\ref{fig1}(a)]


\begin{equation}\label{bose-hubbard-space}
\begin{split}
\frac{\hat{H}}{\hbar} =\; & \omega_{\rm r} \sum_{i=1}^N  \hat{a}_i^{\dagger} \hat{a}_i
- J \sum_{\langle i,j \rangle}^N  \hat{a}_i^{\dagger} \hat{a}_j  - \frac{U_{\rm r}}{2}\sum_{i=1}^N  \hat{a}_i^{\dagger 2} \hat{a}_i^2 \\
& 
+ \epsilon \sqrt{U/\gamma} \left( \hat{a}_0 e^{-i\omega_{\rm p}t} + \hat{a}_0^\dagger e^{i\omega_{\rm p}t} \right)
\end{split}
\end{equation}
Here, $\hat{a}_i$ and $\hat{a}_i^{\dagger}$ are the annihilation and creation operators for the $i$-th cavity, $\omega_0$ is the resonant frequency of each cavity. 
The parameter $U_i$ denotes the on-site Kerr nonlinearity, $J$ is the nearest-neighbor coupling strength, $\epsilon$ is the rescaled pump amplitude, and $\omega_p$ is the pump frequency. The decay of each resonator is modeled by a Lindblad dissipator with rate $\gamma$.
The first term describes the energy of each cavity, assuming that they are identical. The second term represents photon hopping between nearest-neighbor cavities. The third term accounts for the on-site Kerr nonlinearity, which suppresses multiple-photon occupation in a single cavity. The fourth term corresponds to the coherent pump, which injects excitations into the first cavity. In addition, the coupling to the environment induces decay, captured in the last term.


Our implementation consists of a chain of 21 nearest-neighbor-coupled, lumped-element, nonlinear resonators~[Fig.~\ref{fig1}(b) false-colored in blue] 
~\cite{scigliuzzo2022, castillo-moreno2025}. For each resonator, an array of 10 Josephson junctions realizes a nonlinear superinductor. The chain is accessed via input and output ports, capacitively coupled to the chain's first and last sites, allowing for direct probing and pumping of the system. 

Our device also includes two frequency-tunable, transmon-type artificial atoms~\cite{koch2007, blais2021}, capacitively coupled to the chain. 
In the following, both transmons are kept detuned from the modes of the resonators (at $\omega_q \approx 3.2~\mathrm{GHz}$), and do not participate in the experiments presented here.

We characterize our Bose-Hubbard chain parameters at low power by measuring the transmission coefficient $\vert S_{21}\vert$. 
The spectrum of the system exhibits a transmission band from 5.018 to 5.857~GHz, with 21 discernible peaks, each reflecting a hybridized mode of the 21 resonators
~[Fig.~\ref{fig1}(c)]. From the response at low power and assuming identical resonators, we extract their bare frequency, $\omega_{\rm r}$, from the center frequency of the band and the nearest-neighbor coupling, $J$, from the transmission bandwidth, $4J$. Comparing this theoretical model with our experimental data, we estimate $\omega_r/2\pi \approx 5.43~\text{GHz}$ and $J/2\pi \approx 209~\text{MHz}$ [App.~\ref{app:deviceParameters}]. In addition, we obtain the total loss of each resonator by fitting the response of the center resonator at low power with the expression $S_{21}(\omega) = 2 \frac{\kappa}{\gamma + 2i\left(\omega - \omega_0 
\right)}
$ \cite{eichler2014}. In this expression, $\kappa$ is the total radiative decay to both the input and output ports, and $\gamma$ is the total decay. We obtain  
$\gamma/2\pi\approx1$~MHz with a non-radiative decay $\gamma_{\rm nr}/2\pi = 10.25$~KHz for mode thirteen, the pumped one in the following experiments. 

\section{Multi-mode phase transition spectroscopy} \label{section:multi-mode_phase_transition_spectroscopy}

\begin{figure}
    \centering
    \includegraphics[width=\linewidth]{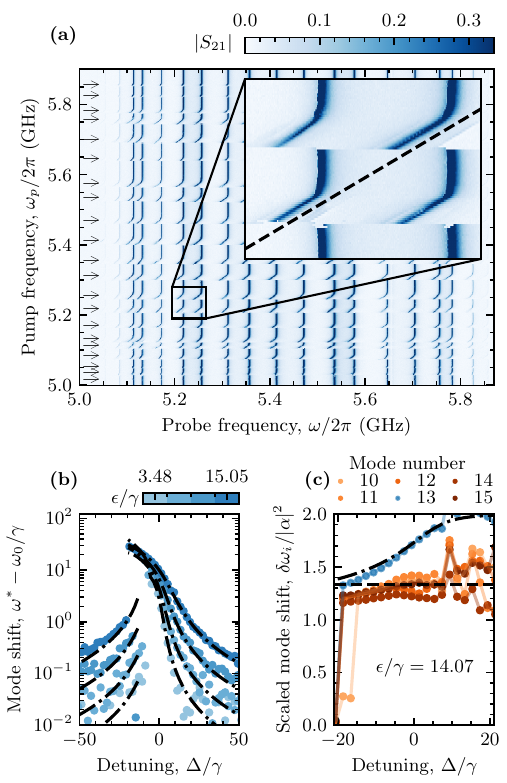}
    \caption{\textbf{Spectroscopic evidence of the phase transition. (a)} Transmission across the chain, $\vert S_{21}\vert$ versus probe and pump frequencies at a fixed pump amplitude $\epsilon/\gamma=14.07$.
    Each horizontal arrow corresponds to the resonant frequency of a mode of the undriven chain. When the pump is close in frequency to any of the modes (dashed black line in the Inset), the resonance frequencies of all modes shift abruptly. \textbf{(b)} Mode-13 frequency shift versus normalized pump detuning, $\Delta/\gamma$ for selected pump amplitudes (color bar). Theory model results are shown by black dashed lines. \textbf{(c)} Normalized frequency shifts of center-band modes (indices 10 to 15), $\delta \omega_i /(\gamma |\alpha|^2)$ versus $\Delta/\gamma$ at fixed pump amplitude. Corresponding theory predictions for non-resonant (resonant) modes are plotted as dashed (dot-dashed) lines.
    }
    \label{fig2}
\end{figure}

We add a coherent pump at a frequency $\omega_{\rm p}$, and measure the transmission $|S_{21}|$ through the metamaterial while sweeping the frequency of the pump at a constant amplitude~[Fig.~\ref{fig2}(a)] (see App.~\ref{section:power_dependence} for pump power dependence).
As the pump frequency approaches the resonant frequency of one of the modes in the chain from above, the mode undergoes a frequency shift up to a few tens of MHz. When the detuning becomes too large, the frequency shift abruptly stops, and the mode returns to its initial state. 
This behavior has previously been observed in other systems~\cite{muppalla2018} and characterizes a first-order phase transition. 

Remarkably, not only does the resonant mode experience the frequency shift, but changes of comparable magnitude are observed across all modes of the array, including those far detuned from the drive frequency.

We propose the following model to account for the observed behavior.
Expressed in terms of Fourier modes of the array, the Bose-Hubbard Hamiltonian in Eq.~\eqref{bose-hubbard-space} reads 
\begin{equation}
    \hat{H} = \sum_{k=1}^N \omega_k \hat{a}^\dag_k \hat{a}_k - \frac{U_{\rm r}}{2N} \sum_{k,p,q=1}^{N} \hat{a}^\dag_{p+k} \hat{a}^\dag_{q-k} \hat{a}_p \hat{a}_q,
\end{equation}
with $\omega_k$ the frequency of mode $k$.
The second term sum contains self-Kerr terms $\propto \hat{a}^{\dag 2}_k \hat{a}^2_k$, cross-Kerr terms $\propto \hat{a}^\dag_q \hat{a}_q \hat{a}^\dag_p \hat{a}_p$ and finally photon conversion terms that we neglect on the basis that they are not energy conserving. 
We postulate that only the quasi-resonant mode---the mode with the smallest detuning from the pump---is effectively driven, acquiring a finite photon population. The other modes are only shifted in frequency by cross-Kerr effect. 

We are left to the study of an effective single-mode bosonic system. We write its Hamiltonian in the pump frame of reference, 
\begin{equation}
    \hat{H} = 
        -\Delta \hat{a}^\dag \hat{a} 
        - \frac{U}{2N}\hat{a}^{\dag 2}\hat{a}^2 
        + \sqrt{\frac{U}{\gamma N}} \epsilon (\hat{a}^\dag + \hat{a}),
\end{equation}
in which $\Delta=\omega_{\rm p} - \omega_0$ is the detuning between the mode bare resonance frequency and the pump.
We add that the mode is subject to single-photon losses, modeled by a Lindblad jump operator $\hat{L}=\sqrt{2\gamma}\hat{a}$, with $\gamma$ the total loss rate. 
We 
define an effective pump amplitude, $\epsilon$, that is rescaled by the interaction strength.

The steady-state of this driven-dissipative system is well known for its bistability~\cite{drummond1980}: in a certain $\epsilon$ and $\Delta$ parameter regime, the system has two stable semi-classical steady states, called bright and dim states based on their respective high and low population number. 
The quantum steady-state, on the other hand, must be unique~\cite{nigro_uniqueness_2019}: if the pump amplitude is raised, it rapidly goes from dim to bright. Finally, when $U/(\gamma N) \to 0$, the photon population becomes macroscopically large. The quantum fluctuations consequently become smaller, and mean-field predictions are more accurate. It can be interpreted as a thermodynamic limit due to the large number of particles in interaction; this intuitive picture is made rigorous by a path integral computation~\cite{sepulcre2025}. Finally, when fluctuations vanish entirely, the limit between the dim and bright regimes becomes a sharp, first-order phase transition.
In our case, the effective nonlinearity is further reduced by the presence of multiple Josephson junctions and scales as $U/(\gamma N N_{JJ}^2)$~\cite{pechal2016}. With this scaling factor, our system reaches $U/(\gamma N N_{JJ}^2) = 0.123 $.

We observe the sharp jump in photon occupancy by the renormalization of the mode resonance frequency. To compute analytically this frequency renormalization, we first obtain the mode response functions in their steady-state, given by Gaussian fluctuations around the mean-field value of the Keldysh partition function. 
Then, we locate the pole of the response function [App.~\ref{app:theoreticalModel}], at 
\begin{equation}
    \omega_* = \sqrt{(\Delta+2\gamma|\alpha|^2)^2 - \gamma^2|\alpha|^4} + \omega_{\rm p},
\end{equation}\label{eq:self_kerr}
with $\alpha = \sqrt{U/\gamma N}\avg{\hat{a}}_{\rm ss} $ the mean-field value of the photon annihilation operator. Then, $\alpha$ is obtained by solving the mean-field self-consistent equation
\begin{equation} \label{eq:selfConsistent}
    \left(1 + \left(\frac{\Delta}{\gamma} + |\alpha|^2 \right)^2\right)|\alpha|^2 = \frac{\epsilon^2}{\gamma^2}.
\end{equation}


We compare the model to spectroscopy measurements of a single mode under varying pump powers. The global scale of $\epsilon$ is treated as a single fit parameter, 
and adjusted to achieve excellent agreement between frequency shift measurements and the model~[Fig.~\ref{fig2}(b)]. 
Notably, this single fit parameter suffices to capture all the pump power values, on both sides of the transition, across every mode of the chain. 

Finally, we confirm that the cross-Kerr coefficients are responsible for the frequency shifts by comparing the ratio between the frequency shifts of several modes to\footnote{We assume that all mode have approximately the same losses \(\gamma\), and \(\alpha=0\) for all modes but the quasi-resonant one, since they are far detuned from the drive.} $\gamma |\alpha|^2$. Numbering the modes from lowest frequency (1) to highest (21), we measure the central modes, 10 to 15, since they exhibit more visibility in the spectroscopy measurements, while pumping mode 13. 

A detailed computation~[App.~\ref{app:theoreticalModel}] in the case of closed boundary conditions predicts a ratio of $4/3$ for all modes, which is coherent with experimental evidence~[Fig.~\ref{fig2}(c)]. The result is consistent for all modes, even if the mode frequencies deviate from the ideal equal resonator model, due to disorder in the resonators coming from the Josephson junctions~\cite{scigliuzzo2022, weissl2015}. Since the pump scaling has been calibrated on the previous measurement, this fit does not require any free parameters. Note that, as expected, the quasi-resonant mode (in blue), due to self-Kerr, follows Eq.~\ref{eq:selfConsistent} instead and thus undergoes a different shift, and does not align with the other modes, shifted by cross-Kerr.

\section{Phase diagram} \label{section:phase_diagram}

While spectroscopy measurements provide information about the frequency shifts and quantitative experiment-to-model agreement, it is important to underline that the abrupt jump we observe in the frequency shift does not align with the transition line, predicted, \textit{e.g.}, by numerical exact diagonalization of the Liouvillian of the system. 

Indeed, close to the transition, bistability manifests itself by hysteresis~\cite{rodriguez2017}: past the transition, the system stays trapped in its initial state, now metastable, for a macroscopically long time scale. 
The sudden jumps we observe then correspond to the border of the bistability region, where the metastable state vanishes and the system falls back into its true ground state~[App.~\ref{section:Hysteresis}]. 

\begin{figure*}
    \centering
    \includegraphics{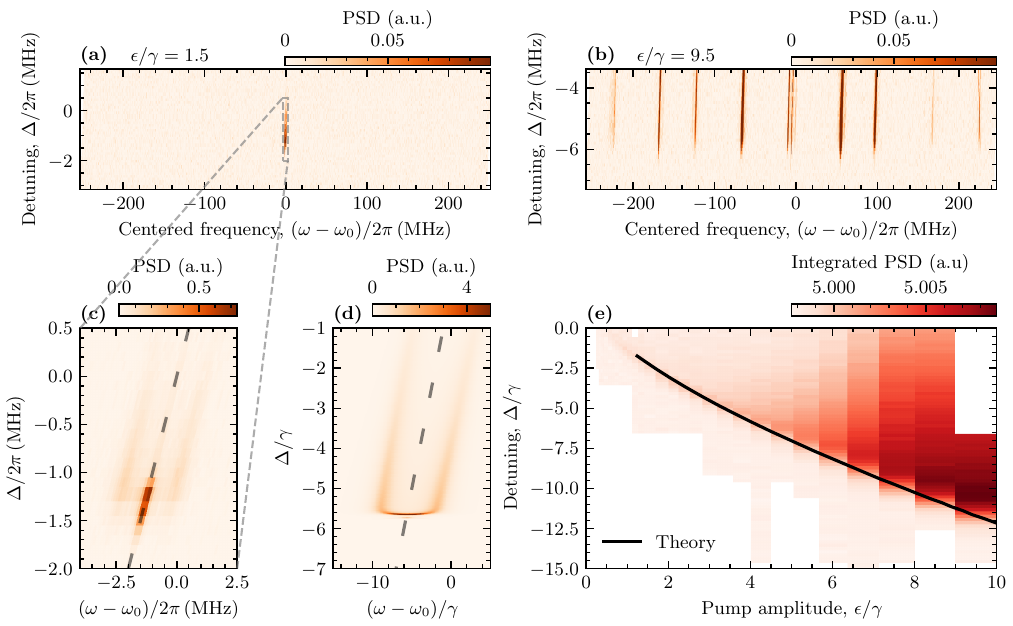}
    \caption{\textbf{Photon emission from the metamaterial. (a)} Power spectral density (PSD) against detuning over several modes at moderate pump power $\epsilon/\gamma=1.5$. Only the quasi-resonant mode emits. \textbf{(b)} Power spectral density (PSD) against detuning over several modes at a larger pump power, $\epsilon/\gamma=9.5$. Other modes of the chain start to emit. \textbf{(c)} Zoom on the resonant mode emission spectrum. Theory prediction of the emission spectrum. \textbf{(d)} Theory prediction of the emission spectrum in (c). \textbf{(e)} Integrated power spectral density against detuning and pump amplitude. High emission indicates a high photon number stored in the mode: we observe a phase boundary between the dim and bright states. The numerically computed transition line is in black.}
    \label{fig3}
\end{figure*}

To accurately determine the transition line, we measure the power spectral density (PSD) of the system as a function of the pump frequency and amplitude~[Fig.~\ref{fig3}(a),(b),(c)]. For this measurement, the probe signal, used in the spectroscopy experiments, is switched off. We sweep the pump frequency and power, directly recording the emission from the chain.
Furthermore, the measurements are performed in an interleaved sequence (pump on / pump off) to suppress the effect of thermal excitations and fluctuations. Residual pump leakage is subsequently removed during post-processing through signal filtering.

We observe two distinct emission regimes depending on the pump power. For moderate pump power, $\epsilon/\gamma=1.5$~[Fig.~\ref{fig3}(a)], the emission occurs at a single (mode) frequency. However, as the power increases and $\epsilon/\gamma=9.5$~[Fig.~\ref{fig3}(b)], additional emission peaks emerge at other mode frequencies of the chain. Our model, based on a single-mode excitation, captures only the behavior for single-mode emission.

A closer inspection around the transition~[Fig.~\ref{fig3}(c)] reveals clear changes in spectral structure. Far detuned from the transition, the system remains in the vacuum state, and no emission is detected.
At the transition, a strong emission peak appears at the pump frequency. Then, deeper into the bright side of the transition, we detect two emission peaks: one at the renormalized resonator bright frequency, and another one symmetrically located with respect to the pump frequency.
This double peak is a hallmark of the strongly nonlinear effects in the bright phase dominated by the self-Kerr energy term; it is similar in nature to a Mollow triplet~\cite{mollow1969}.

We find a good qualitative agreement between our measurement and the model result~[Fig.~\ref{fig3}(d)]. The incoherent emission spectrum is computed from the Keldysh response function~[App.~\ref{app:theoreticalModel}].
The obtained analytical expression produces two emission peaks when $\omega_*>\gamma$, which in practice is verified everywhere but at the transition. 
Here, we obtain a single strong emission peak around $\omega_{\rm p}$. 
In this case, the mean-field value is obtained numerically by the exact diagonalization of the Liouvillian in the $U/(\gamma N)\to0$ limit. 
The pronounced emission peak coincides with the abrupt jump of $\alpha$
between its dim and bright values, as predicted by
Eq.~\eqref{eq:selfConsistent}.

From the measured PSD, we extract the emitted photon number, $\langle \hat{a}^{\dagger}\hat{a}\rangle$, by integrating the emitted signal. 
Repeating this photon-number procedure over a range of pump frequencies and powers directly maps out the phase-diagram of the system~[Fig.\ref{fig3}(e)]. Increasing the pump power shifts the dim-to-bright phase transition to progressively lower pump frequencies, while also extending the lifetime of the bright state. These trends are consistent with the spectroscopy results~[App.~\ref{section:power_dependence}]. 

The measured phase diagram can be compared to a numerical transition line obtained from the single-mode model, 
with the pump amplitude scale as the sole free parameter. Adjusting this single parameter produces excellent agreement between experiment and theory. The PSD measurements exhibit a difference of 
$\qty{4}{\deci\belm}$ 
less power compared to spectroscopy, which is coherent with the experimental setup, 
as we remove the probe signal~[App.~\ref{appendix:probe}].

\section{Dynamics close to the phase transition} \label{section:dynamics_close_to_the_phase_transition}

We access the dynamics of the Bose-Hubbard chain at the transition with time-resolved measurements and use them as another experimental tool to independently confirm the location of the phase boundary. 

In a conventional thermodynamic first-order transition, such as the liquid-vapor transition, the two phases coexist in space, manifested as gas bubbles within the liquid or liquid droplets suspended in the gas. A similar phenomenon takes place for this dim to bright phase transition, albeit in time rather than in space: the system will switch between the two metastable steady-states, with a characteristic time exponentially large in the thermodynamic limit, as the nonlinearity $U/(\gamma N)\to0$.  

\begin{figure}
    \centering
    \includegraphics{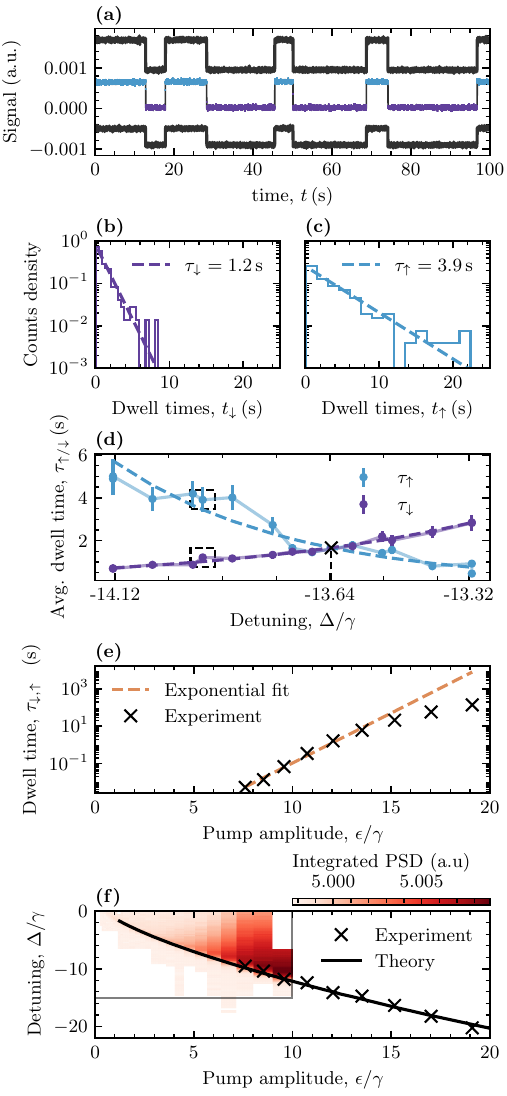}
    \caption{\textbf{Time-resolved measurements. (a)} Portion of measured time trace for the resonant (color) and adjacent modes (gray). The bright dwell times are in purple, the dim dwell times in blue. \textbf{(b)} and \textbf{(c)} Histograms of dwell times reported for one value of detuning and pump amplitude (Square in (d)). Estimated exponential density probability functions in dashes, providing the characteristic dwell times. \textbf{(d)} Characteristic dwell times in bright and dim states against detuning at finite pump amplitude. Their intersection provides the transition location. Exponential model fits in dashes. \textbf{(e)} Dwell times at the first-order transition line. Dashed line: Exponential fit to the data. 
    \textbf{(f)} Extended phase diagram in the detuning/pump amplitude plane. The integrated PSD data within the grey rectangle is the same as in Fig.\ref{fig3}(e). Crosses denote direct measurements of the first-order transition line obtained by dwell-time measurements as in panel (d). The transition line predicted by theory is shown as a black line.}
    \label{fig4}
\end{figure}

We perform a two-tone measurement at the dim-state frequencies while applying a pump around the frequency of mode thirteen. 
We acquire a time-resolved trace over two hours with an integration time of $\qty{0.01}{\second}$ for the four closest modes to the pumped one~[Fig.~\ref{fig4}(a)]. In our trace, the dim state produces a high-signal trace (blue). When the system transitions to the bright state, the trace switches to a low-signal level (purple). We demonstrate that the sudden switches of resonance frequency across the chain’s modes occur synchronously, with all traces exhibiting simultaneous jumps. In the following, we track only the pumped mode.

We obtain the dwell times of the transition by assigning each data point as belonging to the dim or bright states using a threshold set at half the amplitude difference between the two states~[Fig.~\ref{fig4}(b) and (c)].  
Assuming the jumps follow Poisson statistics, the dwell times in the bright $t_\downarrow$ and dim $t_\uparrow$ states must follow an exponential distribution, $P(t) = 1/\tau \exp (-t/\tau)$, with characteristic dwell times $\tau_\uparrow$ and $\tau_\downarrow$, respectively. The maximum likelihood estimator for $\tau$ is simply the average of the sample of dwell times. We obtain the dim and bright characteristic times and observe a good agreement between the exponential fits and the measured dwell-time histograms.

As superconducting circuits are subject to several sources of noise occurring on long time scales, such as $1/f$ noise sources, slow drift in the dilution fridge temperature, or even cosmic ray events, and could significantly perturb a measurement over such a long acquisition time. 
To detect possible contamination of the statistical samples, we conducted an Anderson-Darling test \cite{anderson1952} and rejected time traces failing at $15\%$. 

By repeating the protocol for various detuning values at a fixed pump amplitude, we monitor the jump rates across the bistability region~[Fig.~\ref{fig4}(d)]. These rates are expected to obey a generalized Boltzmann form, $1/\tau \propto \exp(-E(\Delta, \epsilon) / U)$, where $E$ represents an activation barrier and $U$ measures the strength of the quantum fluctuations triggering the jumps~\cite{sepulcre2025}. 
Assuming $E(\Delta, \epsilon)$ is linear in the vicinity of the transition line, we propose an exponential fit of the switching rates (dashed lines).  The intersection of the two fits identifies the detuning at which the dim and bright states are equiprobable, marking the transition point.

We repeat this measurement at different pump amplitudes. The obtained time scales at the transition line range from a few $\unit{\milli\second}$ to $\qty{143}{\second}$~[Fig.~\ref{fig4}(e)], prompting adjustments to both the recording and integration times. Each time trace is required to contain over one hundred switching events, with integration times varied between $\qty{1}{\milli\second}$ and $\qty{0.1}{\second}$ accordingly. 

The dwell times exhibit distinct trends depending on the pump amplitude. For $\epsilon/\gamma < 9.5$, the times increase exponentially with $e^{1.33\epsilon/\gamma}$. This exponential increase is expected in a first-order phase transition activated by tunneling~\cite{risken1987}. Above this threshold, the dynamics saturate, which we attribute to either changes in the activation mechanism or to other sources of error at these slow time scales. The longest timescale measured is 143~s, exceeding values reported in previous studies in other setups~\cite{berdou2023}.

Finally, we report the transition points on the phase diagram in the $(\Delta/\gamma, \epsilon/\gamma)$ plane~[Fig.~\ref{fig4}(f)]. Since the pump amplitude has already been calibrated for probed and non-probed measurements, there are no additional parameters to fit. We report an excellent agreement between the transition lines predicted by theory and measured via jump rate statistics. 
This measure also supplements the one provided by PSD measurements, and extends our measurement of the transition line to larger detunings and pump amplitudes.

We expect that the point where the line of first-order transition stops displays critical behavior: bistability should disappear with switch times reaching \(\tau\to0\), while the response function characteristic time diverges [App.~\ref{app:theoreticalModel}]. This region is currently only accessible to PSD measurements, and out of reach of the time-resolved approach. Narrowing down its location by lowering \(U/(\gamma N)\) (thus raising switching times to measurable values) could be the aim of follow-up investigations.

\section{Discussion and conclusion} \label{section:conclussion}

In this paper, we reported the first experimental observation of multimode dissipative first-order phase transitions in a driven 1D Bose–Hubbard chain comprising 21 nonlinear resonators. Transmission measurements revealed that pumping any single mode induces a uniform frequency shift across all modes, a behavior reproduced by a minimal single-mode mean-field model and attributed to cross-Kerr interactions. Power spectral density (PSD) measurements characterized the dim-to-bright phase transition and, if integrated over pump power and frequency, yielded a phase diagram in quantitative agreement with theory at moderate powers.  
Time-resolved measurements revealed switching rates orders of magnitude below any time scale of the system, up to 143 s, 
and located the transition point via equal lifetimes of the bright and dim states, with the resulting transition line matching both theory and PSD data. At high pump powers, we observed an additional multimode emission not captured by our current model. 

Our results represent a significant step forward in understanding quantum phase transitions and many-body dynamics in multimode photonic systems. An important direction for future work is to extend the theoretical model to include the collective processes that emerge at high powers, in which non-energy-conserving terms are expected to become relevant. Furthermore, we observed no clear feature of the critical point, which could be addressed in further studies. The demonstrated platform functions as a collective switch, with 21 individual modes coherently controlled by a single global drive.  Building on earlier proposals, we identify this platform as a promising candidate for applications in quantum metrology and sensing.


\section{Acknowledgments}

The authors thank Linus Andersson for fabricating the sample package and Riccardo Borgani, Mats Tholén, and Simon Sundelin for the assistance during interleaved PSD measurements. The device in this work was fabricated in Myfab, Chalmers, a micro and nanofabrication laboratory, and measured using Intermodulation Products: Presto. This work received support from the Swedish Research Council; and the Knut and Alice Wallenberg Foundation through the Wallenberg Center for Quantum Technology (WACQT).
T.H. acknowledges financial support from the Chalmers Excellence Initiative Nano.
S. G. acknowledges financial support from the European Research Council via Grant No. 101041744 ESQuAT.

\appendix
\section{Experimental setup}

\begin{figure}[h]
    \centering
    \includegraphics[width=\linewidth]{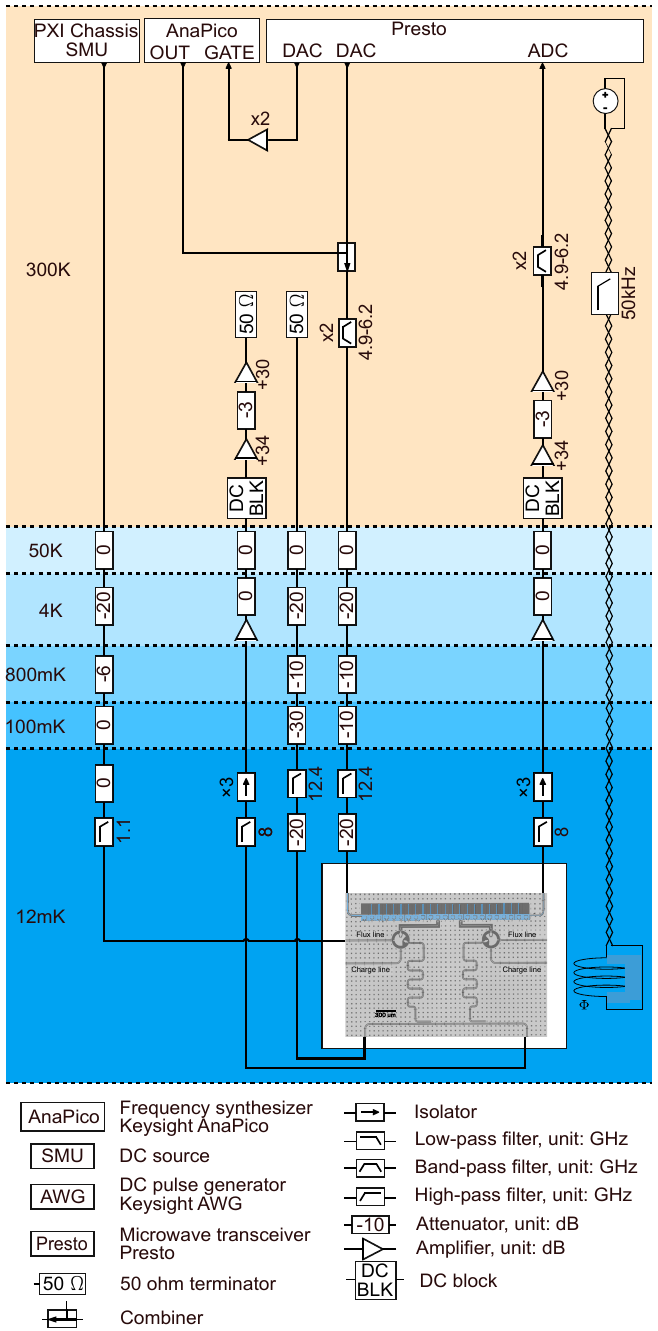}
    \caption{\textbf{Fridge connections.
    } Different thermal stages are color-coded by temperature. Room temperature equipment is at the top, with signal lines leading to the device, including attenuators, amplifiers, and filters to enhance signal-to-noise ratio (SNR). Legend is at the right.}
    \label{fridge}

\end{figure}

Figure~\ref{fridge} illustrates the full wiring diagram of our experimental setup, along with a legend identifying the various instruments and microwave components involved. The device is mounted inside a dilution refrigerator at 12 mK. To protect the sample from electromagnetic and magnetic disturbances, it is enclosed in an RF-tight copper shield and $\mu-$metal enclosure, respectively. Incoming signals are heavily attenuated to suppress thermal photon noise, with 0 dB attenuators as thermalization stages along the coaxial lines. On the output side, the signal is amplified through several stages, starting with a cryogenic High Electron Mobility Transistor (HEMT) amplifier followed by two room-temperature amplifiers. To reduce noise outside the target frequency band and suppress aliasing, we add band-pass filters at room temperature and low-pass filters at low temperature.


We use a microwave transceiver (Presto, from Intermodulation Products AB) capable of direct digital synthesis in the band of interest. One of Presto's output channels is used to gate the output signal from a Keysight AnapiCo frequency synthesizer, the pump. The other Presto's output channel probes the frequency modes of our sample. Both signals are combined with a combiner at the input of the metamaterial. We send flux DC signals with a Source Measure Unit (SMU) from our Keysight PXI chassis through the microwave lines and a coil in a twisted-pair DC configuration. DC flux keeps the emitters detuned from the transmission band.

\section{Superconducting circuit sample}

\begin{figure}
    \centering
    \includegraphics[width=\linewidth]{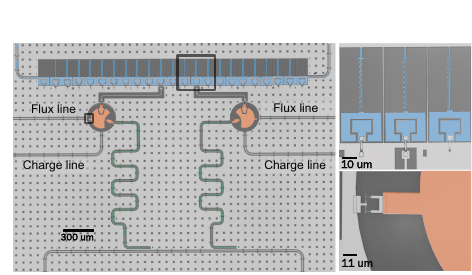}
    \caption{\textbf{False-colored image of the device.
    } Connections to the transmon are labeled for clarity. In blue: Resonator array, in orange: Two transmon qubits, in green: Readout resonators for the transmons. Insets outlined in black: Top corresponds to a close-up of three resonators with the array of 10 Josephson junctions. Bottom: Asymmetric SQUID for the transmon.}
    \label{sample}
\end{figure}

Our complete superconducting circuit sample in~Fig.~\ref{sample}, features a Bose-Hubbard chain composed of 21 high-impedance microwave cavities. 
Each cavity has an inductor consisting on 10 Josephson junctions connected in series and a capacitor created by a metal pad connected to ground. The Josephsons junctions provide large inductance, resulting in a high characteristic impedance~$Z_r\approx390~\Omega$ and, consequently, a strong coupling between adjacent cavities. 
Coupling strength is set by the geometry and physical spacing of adjacent cavities and it is denoted by $J$. The total transmission bandwidth of the metamaterial spans $4J$~\cite{castillo-moreno2025}.

Although the sample includes two identical transmon qubits capacitively coupled near the center of the metamaterial (sites 10 and 13), they were not used in the present measurements. These transmons use asymmetric superconducting quantum interference devices (SQUIDs), which gives rise to two flux-insensitive points ("sweet spots"). For the present study, however, the transmons were deliberately biased far away from the Bose–Hubbard resonances, so they do not participate in the measurements. 
To further improve uniformity and symmetry across the array, shadow inner pads connected with airbridges to ground are added within each cavity, reducing fabrication-induced mismatches.

\section{Extraction of device parameters} \label{app:deviceParameters}

Table~\ref{characterization_table} lists the parameters extracted from the characterization measurements and microwave simulations of the Bose-Hubbard chain. At low powers, we can approximate our system to a tight-binding model. Assuming identical resonators, the Hamiltonian given by
\begin{equation}
\begin{aligned}
H & =\sum_{n=1}^N \omega_r a_n^{\dagger} a_n - \sum_{n=1}^{N-1} J \left(a_n^{\dagger} a_{n+1} + a_{n+1}^{\dagger} a_{n} \right),
\end{aligned}
\label{eq:tightbinding}
\end{equation}
In this model, $\omega_r$ is the bare resonator frequency, $J$ is the nearest-neighbor coupling between resonators, $a_n$ ($a_n^{\dagger}$) is the photon annihilation (creation) operator of the $n$-th resonator. Both $\omega_{r}$, $J$ are obtained by comparing the Hamiltonian with the transmission measurement in our metamaterial in~Fig.~\ref{fig1}(c), as reported in the main text. 
These values are supported by microwave simulations of the system and room temperature measurements of the resistance of the Josephson junctions, from which we extract the capacitances and inductances, respectively. Microwave simulations yield consistent values, $ \omega_{r}/2\pi \approx 5.37 $~GHz and $ J/2\pi\approx 190 $~MHz, with the small discrepancies attributed to disorder in the Josephson junction arrays.

In addition, we extract the total decay, $\gamma_T$, and the coupling to ports, $\kappa$, from a Lorentzian fit to the transmission of the resonator, as reported in the main text. The impedance of the resonators is extracted from  $Z_r = 1/\omega_r C_r$, with $C_r$ the resonator capacitance from simulations. With this capacitance value, we extract the scaled Kerr-nonlinearity as $U^\prime=\frac{e^2 }{2C_rN_{JJ}^2N_r} = 123$~KHz~\cite{pechal2016, masluk2012}.

\begin{table}
    \centering
    \caption{Model parameters and experimentally determined values.}
    \begin{ruledtabular}
    \begin{tabular}{lcc}
         \textbf{Parameter} & \textbf{Symbol} & \textbf{Value} \\ \hline \hline
        Metamaterial center frequency & $\omega_r/2\pi$ & 5.43 GHz \\
        Nearest-neighbor coupling & $J/2\pi$ & 209.7 MHz \\ 
        Scaled Kerr-non linearity & $U/(2\pi N N_{JJ}^2)$ & 123 KHz \\
        Resonator radiative decay & $\kappa/2\pi$ & 1 MHz  \\ 
        Resonator non-radiative decay & $\gamma_{nr}/2\pi$ & 10.25 KHz  \\ 
        Resonator impedance & $Z_r$ & 387.23 $\Omega$\\
    \end{tabular}
    \end{ruledtabular}
     
\label{characterization_table} 
\end{table}

\section{Theoretical model} \label{app:theoreticalModel}
\subsection{Single-mode model}
We focus on a single Fourier mode of the array. The Hamiltonian governing the dynamics of the mode, as written in the main text, is 
\begin{equation}
    	\hat{H} = 
		- \Delta \hat{a}^{\dagger}\hat{a} 
		- \frac{U}{2N} \hat{a}^{\dagger2} \hat{a}^2
		+ \sqrt{\frac{\gamma N}{U}}\epsilon(\hat{a}^{\dagger} + \hat{a}).
\end{equation}
Note that we expressed it in the reference frame of the drive, such that $\Delta = \omega_{\rm p} - \omega_{0}$. We also preemptively rescaled the pump amplitude $\epsilon$ by the Kerr interaction coefficient, such that it stays constant in the limit $U/(\gamma N)\to0$, which corresponds to a thermodynamic limit where the photon occupation number diverges. We also include single photon losses in the model in the form of a Lindblad jump operator $\hat{L}=\sqrt{2\gamma}\hat{a}$, where $\gamma$ accounts for all loss channels. 

To deal with driven-dissipative systems out of equilibrium, it is convenient to use the Keldysh path integral representation~\cite{sieberer2016}. 
This allows us to perform a self-consistent mean-field approximation and derive the response functions of the system around this mean-field solution in one, coherent framework. The main object of the theory is the partition function, $Z=\int\mathfrak{D}[\alpha_c, \alpha_q, \alpha_c^*, \alpha_q^*] e^{iS}$, with $\alpha_c$, $\alpha_q$ the classical and quantum fields defined from the forward and backward fields ($\alpha_+$ and $\alpha_-$ resp.), by $\alpha_c = (\alpha_+ + \alpha_-)/\sqrt{2}$ and $\alpha_c = (\alpha_+ - \alpha_-)/\sqrt{2}$. 
The Keldysh action associated with our model is 
\begin{align}
    	S = \int{\rm d}t\, \alpha^{\dagger} \mathbf{D}(t) \alpha 
	&+ \frac{U}{2}(|\alpha_c|^2 + |\alpha_q|^2)(\alpha_c^*\alpha_q + \mathrm{h.c.}) \nonumber\\ 
	&- \sqrt{\frac{2 \gamma N}{U}} (\epsilon \alpha_q^* + \epsilon^* \alpha_q),
\end{align}
where we employed a vector notation, $\alpha = (\alpha_c, \alpha_q)$ and $\mathbf{D}(t)$ is a matrix of advanced, retarded, and Keldysh propagators, 
\begin{equation}
    	\mathbf{D} =
	\begin{bmatrix}
     0 & D_{\mathrm{A}}  \\
     D_{\mathrm{R}} & D_{\mathrm{K}}
	\end{bmatrix} = 
	\begin{bmatrix}
     0 & i \partial_t + \Delta -i \gamma  \\
     i \partial_t + \Delta + i \gamma & 2i \gamma
	\end{bmatrix}.
\end{equation}

\paragraph*{Mean-field.} The mean-field equation is obtained by decomposing the classical field into a mean value and fluctuations, $\alpha_c(t) = \bar\alpha_c + \delta\alpha_c(t)$. The quantum component cannot acquire a mean-field value; in the sake of consistency, we denote it $\alpha_q(t) = \delta\alpha_q(t)$. Injecting it in the action, we develop and collect all linear terms in $\delta\alpha_c$, which read
\begin{equation} \label{eq:firstOrder}
    	\int{\rm d}t\, \delta \alpha^*_q\left( D_{\mathrm{R}}(t) \bar{\alpha}_{c} - \sqrt{\frac{2\gamma N}{U}} \epsilon + U\frac{|\bar{\alpha}_c|^2}{2} \bar{\alpha}_c \right) + \mathrm{h.c.}
\end{equation}
The bracketed term corresponds to an effective drive field imposed on the fluctuations. We impose the self-consistent condition $\avg{\delta\alpha_c}=0$, meaning that $\delta\alpha_c$ is a pure fluctuation field with no average value. Truncating the action at quadratic order, this simply imposes that the effective drive term is zero. We rearrange terms and square the conditions to obtain the mean-field self-consistent equation from the main text, 
\begin{equation}
    \left(1 + \left(\frac{\Delta}{\gamma} + |\alpha|^2 \right)^2\right)|\alpha|^2 = \frac{\epsilon^2}{\gamma^2}.
\end{equation}
where we have used the shorthand $\alpha = \sqrt{U/ 2 \gamma N} \bar\alpha_c$. This quantity is linked to quantum observables by the relation $\bar\alpha_c= \avg{\alpha_c}=\int\mathfrak{D}[\alpha_c,\alpha_q] \alpha_c e^{iS} = \sqrt{2}\avg{\hat{a}}$.
Since the self-consistent equation is a cubic polynomial in $|\alpha|^2$, several analytical results can be derived from Cardano's method for root finding. 
First, we have a critical point at $\Delta/\gamma = -\sqrt{3}, \epsilon/\gamma=2\sqrt{2}/\sqrt{3\sqrt{3}}$, where the system has a triple root. 
In the phase diagram, this is where the first-order transition line terminates. 
Below the critical point exists a bistable region, where the three distinct roots correspond to the dim and bright state, with an extra unstable solution. 
This region is delimited by the lines where degenerate roots are found, 
\begin{equation}
    	\frac{\epsilon}{\gamma} = 
		\sqrt{-\frac{2\Delta}{3\gamma}\left( 1+\left( \frac{\Delta}{3\gamma} \right)^2 \right) 
		\pm 2\left( \left( \frac{\Delta}{3\gamma} \right)^2 - \frac{1}{3}\right)^{3/2}}.
\end{equation}

\paragraph*{Response functions.} At the thermodynamic limit $U/(N\gamma)\to0$, the mean-field treatment is exact. The ground state is Gaussian, displaced by $\bar\alpha_c$. To compute the response functions, we return to the development of the action in terms of fluctuation fields $\delta\alpha_c$ and $\delta\alpha_q$. After the first-order in \(\delta\alpha_q\) [Eq.~\eqref{eq:firstOrder}], we collect quadratic terms. We note that anomalous terms have appeared, of the form $\delta \alpha \delta \alpha$ or $\delta \alpha^* \delta \alpha^*$. We extend the formalism using Nambu bosons~\cite{sieberer2016}, $\mathfrak{a}(\omega) = (\alpha_c(\omega), \alpha_c^*(-\omega), \alpha_q(\omega), \alpha_q^*(-\omega))/\sqrt{2}$, and the quadratic part of the action writes 
\begin{equation}
    	S = \int{\rm d}t\, \mathfrak{a}^{\dagger} \mathbf{D}'(t) \mathfrak{a}, \quad
	\mathbf{D}' = 
	\begin{bmatrix}
		0 & \mathbf{D}'_{\mathrm{A}} \\
        \mathbf{D}'_{\mathrm{R}}  & \mathbf{D}'_{\mathrm{K}}
	\end{bmatrix},
\end{equation}
each block being itself a $2\times2$ matrix. The retarded block is 
\begin{equation}
    \mathbf{D}'_{\mathrm{R}} = 
	\begin{bmatrix}
		i\partial_t + \Delta + i\gamma + 2\gamma |\alpha|^2 & \gamma\alpha^2 \\
        \gamma\alpha^{*2} & -i\partial_t - \Delta - i\gamma + 2\gamma|\alpha|^2 
	\end{bmatrix}.
\end{equation}
We move to the frequency domain, and compute the response functions from the propagators as 
$\mathbf{G}_{\mathrm{R}}(\omega) = \mathbf{D}_{\mathrm{R}}^{\prime-1}(\omega)$ and $\mathbf{G}_{\mathrm{K}}(\omega) = - \mathbf{G}_{\mathrm{R}}(\omega) \mathbf{D}'_{\mathrm{K}}(\omega) \mathbf{G}_{\mathrm{A}}(\omega)$. Consequently, the poles of the retarded functions are found by solving $\det\mathbf{D}'(\omega)=0$. Their expressions are
\begin{equation}
    \omega_{\pm} = \pm \sqrt{(\Delta + 2\gamma|\alpha|^2)^2 - \gamma^2|\alpha|^4} - i\gamma.
\end{equation}
Several cases must now be distinguished. 

If $\Delta > -|\alpha|^2$, the square root is real. The imaginary part, linked to dissipation, is unaffected by the build-up of photon population, while the real part, the resonance frequency, shifts. The system dissipates energy close to the positive frequency pole $\omega > \omega_{\rm p}$ in the laboratory frame, while the second pole, only present at $|\alpha|^2 > 0$, amplify incoming signals (the imaginary part of the retarded Green's function is positive). 

If the square root is purely imaginary, the response functions present a single peak at zero frequency (or at $\omega_{\rm p}$ in the laboratory frame). Furthermore, at the critical point, since $|\alpha|^2 = -2\Delta/3$, one of the poles is at $\omega=0$. The response functions develop a singularity, $G(\omega) \propto 1/\omega$, and consequently $G(t) \propto \log t$. This is the expected critical slowing down of the system's dynamics. Observing this behavior, checking the location of the critical point, and experimentally verifying the logarithmic response time would be a stimulating experimental challenge, which, to our knowledge, has not been achieved yet.

\paragraph*{Emission spectrum.} Within the Keldysh formalism, this quantity is given by~\cite{zhang2021} $S(\omega) = i\frac{\gamma}{2} (G_{\rm K}(\omega) - (G_{\rm R}(\omega) - G_{\rm A}(\omega)))$. Note that since the response functions only describe the field fluctuations, we only compute the incoherent part of the emission spectrum. Using the shorthand $\omega_\pm = \pm \omega_* - i\gamma$, we write the spectrum as 
\begin{equation}
    	S_{\mathrm{incoh}}(\omega) = \frac{2 |\alpha|^4}{
		\left(1+\left( \frac{\omega - \omega_*}{\gamma} \right)^2\right)
		\left(1+\left( \frac{\omega + \omega_*}{\gamma} \right)^2\right)
	}
\end{equation}
Concerning the dim state, $|\alpha|$ is weak, and no emission is observed. In the bright phase, two emission peaks appear, at $\omega_{\rm p} \pm \omega_*$ in the laboratory frame. They correspond to the amplification and attenuation peaks of the retarded response. Finally, at criticality, a single, strong emission peak is observed at $\omega_{\rm p}$. Elsewhere along the first-order transition line, the value of $|\alpha|$ rapidly grows from its dim state value to its bright state value. In between, it also produces a single, strong emission peak.  

\subsection{Multi-mode model}
We now return to the full Bose-Hubbard Hamiltonian, 
\begin{equation}
    	\hat{H}=\sum_{j=0}^{N-1} 
	\omega_{\rm r} \hat{a}^{\dagger}_j \hat{a}_j 
	- J (\hat{a}^{\dagger}_{j+1} \hat{a}_j + \hat{a}_{j+1} \hat{a}^{\dagger}_j )
	- \frac{U_{\rm r}}{2} \hat{a}^{\dagger 2}_j \hat{a}^2_j.
\end{equation}
We first identify the eigenmodes of the system. This is equivalent to diagonalizing a tri-diagonal matrix. The eigenvectors are given by a discrete sine transform, 
\begin{equation}
    	\hat{a}_k = \sqrt{\frac{2}{N+1}} \sum_{j=1}^N \hat{a}_j \sin\left( \frac{jk\pi}{N+1} \right). 
\end{equation}
The associated orthogonality relation, 
\begin{align}
    	\sum_{j=1}^N \cos\left( \frac{jk\pi}{N+1} \right) &= \begin{cases}
        N, k=2(N+1), \\
        -1 \text{, else,}
        \end{cases}
\end{align}
is used, \textit{e.g.}, to show that the change of variable preserves the commutators, $[\hat{a}_k, \hat{a}^\dag_l]=\delta_{kl}$. The linear part of the Hamiltonian is thus reduced to independent modes, $\sum_{k=1}^N \omega_k \hat{a}^\dag_k \hat{a}_k$, with the dispersion relation $\omega_k = \omega_{\rm r} - 2J \cos{(2\pi k/N)}$. We then insert the solution in the interaction term. Applying the orthogonality relation is cumbersome. We collect all cross-Kerr and self-Kerr terms, which read respectively 
\begin{equation}
    \frac{U_{\rm r}}{N+1} \sum_{k=1}^N  \left(
    \frac{3}{4} \hat{a}^{\dagger2}_k \hat{a}^2_k 
        + \hat{a}^{\dagger}_k \hat{a}_k \hat{a}^{\dagger}_l \hat{a}_l
    \right).
\end{equation}
We identify the mode self-interaction strength at $U=3U_{\rm r}/4$. The effect of the photon population in the quasi-resonant mode, \textit{e.g.} $k=0$, is then a frequency shift of the other modes by $4U/3\avg{\hat{a}^\dag_0 \hat{a}_0} \simeq 4\gamma |\alpha|^2/3$.

\section{Phase transition depending on pump power} \label{section:power_dependence}

\begin{figure}
    \includegraphics[width=0.95\linewidth]{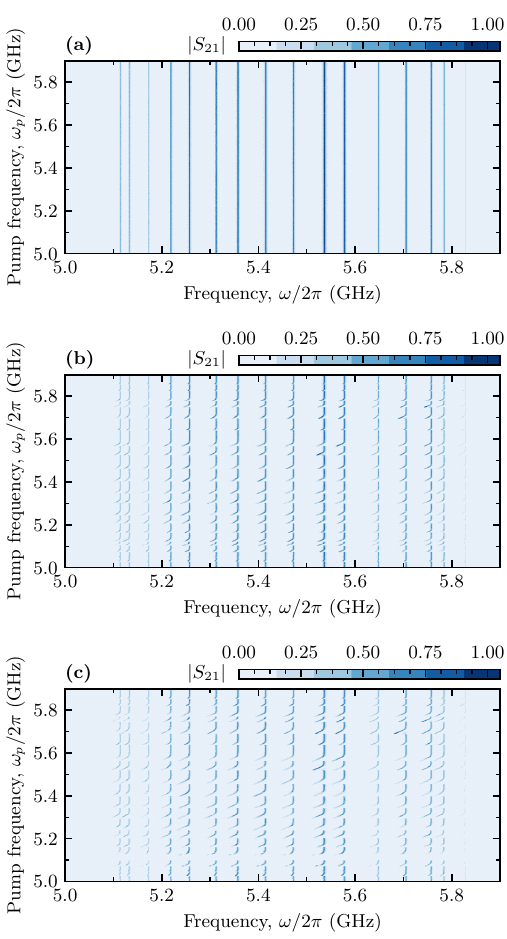}
    \caption{\textbf{Transmission band depending on pump frequency and power. (a)} Low pump power $\epsilon/\gamma=0.14$. The transmission response is not affected by the pump. \textbf{(b)} At intermediate powers, $\epsilon/\gamma=14.07$, first-order transitions appear, reported in the main text. \textbf{(c)} At high pump power $\epsilon/\gamma=28.08$ some modes disappear.}
    \label{jumps_depending_power}
\end{figure}

The response of our system depends strongly on the applied pump power. Probing the complete transmission band, as shown in Fig.~\ref{fig2}(a), while varying the pump power, reveals three different regimes of behavior~[Fig.~\ref{jumps_depending_power}].

At low pump powers, the transmission spectrum remains unchanged across the entire pump frequency range, indicating that the system stays in its dim state with no noticeable interaction between the pump and the resonant modes.

Increasing the pump power leads to the emergence of first-order phase transitions. This is the regime that we study in our main experiments. This multi-frequency response also provides a means to obtain the frequencies of other modes in the chain without directly addressing them.

At even higher pump powers, certain modes disappear from the transmission spectrum for specific pump frequencies. The modes disappearing is a feature of chaos, already reported before~\cite{fitzpatrick2017}.

\section{Hysteresis in the system} \label{section:Hysteresis}

\begin{figure}
    \centering
    \includegraphics[width=\linewidth]{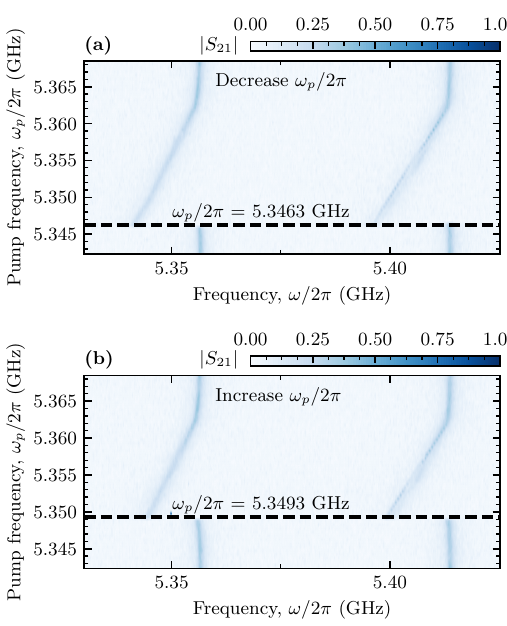}
    \caption{\textbf{Hysteresis of the system. (a)} The pump frequency is swept decreasingly (high to low). \textbf{(b)} The pump frequency is swept increasingly (low to high).}
    \label{hysteresis}
\end{figure}

A hallmark of bistability is the presence of hysteresis around the transition, in which the system’s response to a control parameter depends on its history. The system remains "locked" in its initial phase across a finite range of the control parameter, in this case, the pump frequency, before eventually undergoing a sharp transition to the other phase. 

To prove this behavior, we measure the transmission parameter, $\vert S_{21} \vert$,  while we sweep the pump frequency at a constant power~[Fig.~\ref{hysteresis}]. The sweep is done first from high frequencies to low frequencies~[Fig.~\ref{hysteresis}(a)], in which the system gets trapped in the bright-state, and then from low frequencies to high frequencies[Fig.~\ref{hysteresis}(b)] with the system trapped in the dim-state. The sweep needs to be faster than the jump rates, or the system will relax to the global steady state, and hysteresis is lost.

\section{Probe effect} \label{appendix:probe}

\begin{figure}
    \centering
    \includegraphics[width=\linewidth]{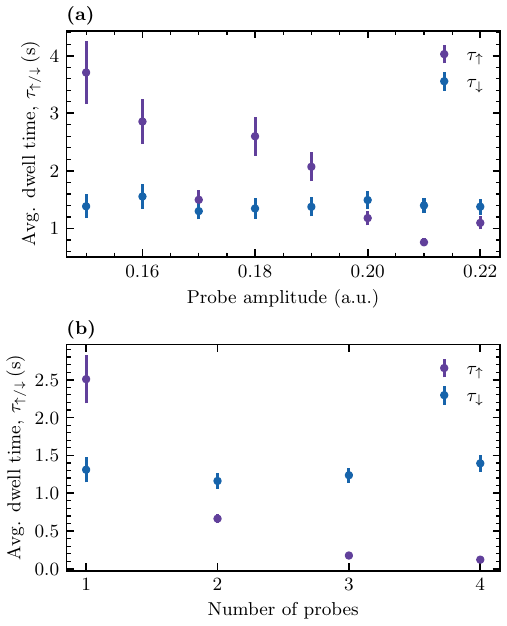}
    \caption{\textbf{Effect of the signal probe on the dynamics of the system. (a)} Effect of the probe amplitude on the rates. \textbf{(b)} Effect of the number of modes being probed.}
    \label{probeeffect}
\end{figure} 

We observe a consistent 4~dBm offset between the transition frequencies measured from the power spectral density (PSD) and those extracted from the jumping rates and or the spectroscopy. We account for this shift in our fitting in~\ref{fig4}(d).

To identify the source of this discrepancy, we study the effect of the probe on the rates~[Fig.~\ref{probeeffect}]. Varying either the probe power or the number of probed modes significantly alters the rates, consistent with earlier observations~\cite{muppalla2018}, and even small probe-power changes (a few photons) have a strong impact. 

Increasing the probe amplitude at fixed pump conditions~[Fig.~\ref{probeeffect}(a)] drives the system from the dim state to the transition point and finally to the bright state, effectively acting as an increase in pump power. Probing additional modes produces a similar effect~[Fig.~\ref{probeeffect}(b)], with each added probe (0.2 a.u.) corresponding to an apparent pump-power increase of $\approx4$dBm, matching the shift caused by our standard single-probe configuration.

\bibliography{Driven-Dissipative_Systems.bib}

\end{document}